\documentstyle[12pt]{article}

\begin{document}

\def \text{\rm}
\def \k{\vec k}
\def \r{\vec r}
\def \hk{\widehat k}
\def \hl{\widehat l}
\def \wh{\widehat}
\def \3{^3{\text He}-{\text A}}
\def \C{\text{C}}
\def \vep{\vec{\partial}}
\def \om{\omega}
\def \be{\beta}
\def \rot{{\text{rot}}}
\def \Dl{\Delta}
\def \dl{\delta}
\def \al{\alpha}
\def \ph{\phi}
\def \cd{\partial}
\def \si{\sigma}
\def \ep{\epsilon}
\def \ga{\gamma}
\def \Ga{\Gamma}
\def \Om{\Omega}
\def \la{\lambda}
\def \IM{{\text{Im}}}
\def \RE{{\text{Re}}}
\def \df{\frac{\text{d}}{\,\text{dx}\,}}
\def \vj{\vec j}
\def \lrot{|\,\hl\times\rot\,\hl\,|}
\def \lll{\matrix{\, \cr ^{\stackrel{<}{\sim}}\cr}}
\def \rrr{\matrix{\, \cr ^{\stackrel{>}{\sim}}\cr}}
\def \BR{I\!\!R}
\def \UN{1\!\!{\text I}}

\title{HIGHER CORRECTIONS TO THE MASS CURRENT IN THE WEAKLY 
INHOMOGENEOUS $^{3}$He -- A}

\author{{\bf C.~Malyshev}\\ \bigskip\\
Russian Academy of Sciences\\
V.A.Steklov Mathematical Institute
\\ St.Petersburg Department
\\Fontanka 27,~St.Petersburg 191011,
{\bf Russia}\\ E-mail:~malyshev@pdmi.ras.ru}

\maketitle
\bigskip

\begin{abstract} 
Two new general representations (the series and the integral) for the mass 
current $\vj$ in weakly inhomogeneous superfluid $A$-phase of Helium--3 are 
obtained near zero of temperature by solving the Dyson--Gorkov equation. These 
representations result in  additional correcting contribution to the standard 
leading expression for $\vj$ which is of first order in gradients of the 
orbital angular momentum vector $\hl$. The total supplementary term 
is found as integral, and, provided the London limit holds, 
the procedure is advanced to expand it at $T=0$ asymptotically by the Laplace 
method in powers of gradients of $\hl$. Three special static orientations of 
$\hl$ with respect to its curl are considered to calculate the higher 
correcting terms up to third order. Coefficients at the quadratic terms are 
estimated numerically, new cubic contributions are found which contain the 
logarithm of the London parameter.
\end{abstract}

\vskip 4truecm
\rightline{PDMI PREPRINT---14/1996}
\rightline{COND-MAT/9612152}
\newpage

\section{INTRODUCTION}

Superfluidity of Helium--3 is firmly in focus of intensive theoretical and 
experimental studies [1]. Although considerable attention is paid last years 
to such problems as quantized vorticity and interfaces [1--4], the weakly 
inhomogeneous A-phase of Helium--3 $(\3)$ also can be a subject for 
theoretical investigation. This phase originates due to $p$-wave spin triplet 
$BCS$--pairing [5,~6], and it demonstrates rather unconventional behaviour 
[7]. Its peculiarity can be seen, for instance, from the mass current $\vj$ 
which is acknowledged to be of first order in gradients [5--7]:
\begin{equation}
\vj_0\,=\,\rho\,\vec v_s\,+\,\frac1{4m}\,\rot\,(\rho\,\hl)\,+\,\vj_{an}
\,,\quad\quad\quad(T=0)
\end{equation}
$$
\vj_{an}\,=\,-\,\frac1{2m}\,{\text C}_0\,\hl\,(\hl\,\cdot\,\rot\,\hl)
$$
where the first two terms are habitual for a nodes--free $p$-wave superfluid, 
while the famous anomalous $\vj_{an}$ witnesses the unusual manifestations 
of the ground state of $\3$. The point is that $\vj_{an}$ is caused by nodes 
existence in the gap on the Fermi surface for real $\3$ [7,~2]. In (1) $\rho$ 
is the liquid density, $m$ is the atom mass, $\vec v_s$ is the superfluid 
velocity, $\hl$ is the weakly inhomogeneous orbital angular momentum vector 
(hat implies a unit vector) and $\C_0 \approx\rho$. The Eq.(1) has been 
deduced by many authors in different approaches: by solving the Gorkov [8--10] 
or the matrix kinetic equations [11], as well as in [12] directly with the use 
of the ground state wave function. 

Alhough the Eq.(1) and the corresponding physical picture have been broadly
discussed [13--21], slight indications can be seen in [8,~12,~20] that higher 
corrections to (1) might occur thus causing difficulties of the superfluid 
hydrodynamics at $T=0$. Indeed, in the course of a phenomenological 
consideration of the free energy of $\3$ Volovik and Mineev have found one of 
these corrections in the form $\chi_{\text{orb}}\,D\,\wh l_a\vep\,\,\wh l_a$, 
where $D=\cd_t+\vec v_s\cdot\vep$ [13]. Aiming to check on their own the 
presence of $\vj_{an}$ in (1), Combescot and Dombre have developed a 
microscopic calculation [10] which has allowed to claim at $T=0$ 
the quadratic correction $\lrot (\rot\,\hl)_{\perp}$ in the current
perpendicular to $\hl$, and the terms $\lrot \left(\vec v_s-(1/4m)\rot\,
\hl\right)_{||}$ and $\lrot(\cd_1\hl_2+\cd_2\hl_1)$ in the current parallel
to $\hl$. 

Alhough [10] has eventually been aimed at the case $T=0$, and the Gorkov 
equation has been solved exactly after linearization of the order parameter, 
an intermediate high-temperature approximation to obtain a manageable formula 
for the $\xi$-integrated Green function has not been avoided. The difference
between the approximate and exact formulas has been considered as responsible
for the second order corrections at $T=0$. However, due to the strategy
adopted in [10], any chance to be accurate with $T\to 0$ and to benefit from 
the exact solution of the governing equation has been lost. Consequently, the 
coefficients at the quadratic corrections and some other formulas have become 
artificially 
complicated, e.g. requiring additional efforts to state that they are finite 
and nonzero. As to the higher contributions at $T=0$, the regular expansion 
procedure has not been convincingly formulated (see discussion in [23]).

Since unambiguous procedure of asymptotic expansion of $\vj$ which would lead
to a more deep knowledge of the ground state of $\3$ seems a meaningful 
technical problem, a way to overcome mathematical
difficulties of [10] has been advanced in [22,~23]. That is another method to 
solve the Dyson--Gorkov equation has been chosen which gave immediately new 
representations for the fermionic Green functions and thus for $\vj$. It is 
well known in mathematical physics that Green function of a Sturm--Liouville 
operator can be written either as integral or as series in eigen--functions. 
The second possibility is meant here. After the subsequent paper [24] it 
became clear that this way looks properly because the new representations 
for $\vj$ admit $T\to 0$ accurately and can rigorously be studied by the 
Laplace method (steapest descent) provided the characteristic length in the 
texture is much longer then the coherence lenght $\xi_0$ (the London limit). 
As the result, it is possible to deduce corrections to (1) systematically
in the form of asymptotic series in powers of gradients of $\hl$. 

This paper completes [22--24] and is organized as follows. Sec.~2 contains 
the outline of the problem which is the same as in [10] (i.e. the 
approximations and notations). Sec.~3 is concerned with the solution of the 
ordinary non-homogeneous differential equation related to the Dyson--Gorkov 
equation and with the calculation of the mass current in the form of series. 
Sec.~4 contains the integral representations for that series as well 
as various limits for the correspondingly written $\vj$: 
lowest ``gradient'' limit and zero temperature limit. 
Three particular cases of mutual orientation of $\rot\,\hl$ and $\hl$ are 
considered in the Sec.~5 to obtain corrections to $\vj_0$ (1) at $T=0$ up to
third order. Apart from the quadratic terms predicted in [10], new cubic
contributions are found which contain the logarithm of the London parameter.
The numerical coefficients at the second order terms are provided. Discussion 
in the Sec.~6 concludes the paper. The present investigation supports 
rigorously the early results of [10,~13] and would be useful for any
systematical microscopic approach to correct observables in $\3$.

\section{THE PROBLEM}

Since our main task is to calculate the mass current $\vj$ by means of normal 
Green function, let us start with the standard matrix Dyson--Gorkov equation:

\def \vk{{\vec k}}
\def \bp{{\prime\prime}}
\def \kbp{{k^\bp}}
\def \vbp{v^\bp}

\begin{equation}
\partial_\tau g(\vk,\vk^\prime)-\int{\text d}^3k^\bp
H(\vk,\vk^\bp)\,g(\vk^\bp,\vk^\prime)
=(2\pi)^3\dl^{(3)}(k-k^\prime)\,\dl(\tau-\tau^\prime).
\end{equation}
Here $\tau$ is ``imaginary'' time, $g(\vk,\vk^\prime)$ is the $2\times2$ 
matrix of normal and anomalous two--point Green functions and $H(\vk,\vk^\bp)$ 
has the form: 
$$
H(\vk,\vk^\bp)=\left(\matrix{
\xi_{k^\bp}\dl^{(3)}(k-\kbp)  &(2\pi)^{-3}\Dl(\vk,\vk^\bp)\cr
(2\pi)^{-3}\Dl^*(\vk^\bp,\vk) &-\xi_\kbp\dl^{(3)}(k-\kbp) \cr}\right),
$$
where $\xi_k\equiv (k^2-k_F^2)/2m$, $k_F$ is the Fermi momentum and
$\Dl(\vk,\vk^\bp)$ is the order parameter of $\3$. We shall calculate $\vj$ by 
the formula
\begin{equation}
\vj\,=\,\be^{-1}\sum\limits_{\om}(2\pi)^{-3}\int{\text d}^3k\,\vk \,g_{11}\,.
\end{equation}
As far as the Refs.[22--24] have been conceived as technical improvement of 
the Ref.[10], the framework (i.e. approximations, notations) turns out to 
be unaffected here, and one should be referred to [10] for certain details.

It is appropriate to re-write (2), (3) in the mixed coordinate--momentum 
representation [12,~25,~26]:
$$
H(\vk,\r)=(2\pi)^{-3}\int{\text d}^3q\,H(\vk+\vec q/2\,,\,\vk-\vec q
/2)\, e^{i\,\vec q\cdot\r}\,,$$
\begin{equation}
g_\vk(\r)=(2\pi)^{-3}\int{\text d}^3q\, g(\vk+\vec q,\vk)\, e^
{i\,\vec q\cdot\r}\,, \end{equation}
where $\r=\frac12\,(\r_1+\r_2)$ is the center of mass coordinate and the 
momentum $\vk$ is conjugated to $\r_1-\r_2$. For instance, it is not difficult 
to check that the relation
$$
(2\pi)^{-3}\int{\text d}^3\kbp\, \Delta(\vk, \vk^\bp)\,
g(\vk^\bp,\vk^\prime)\,=\,
\int{\text d}^3r\, e^{i\,\vec r\cdot(\vk^\prime-\vk)}
\left[\Delta(\vk- i\vep_r-\frac{i}{2}\vep_y\,,\,\vec y)\,g_{\vk^\prime}(\r)
\right]\bigg|_{\vec y=\r}
$$
holds by (4). Applying (4) to (2) we set in the lowest in $\dl/E_{F}$ order:
$$\xi_{(k-i\partial)}\,\approx\,\xi_{k}-i\,m^{-1}\,\vk\cdot\vep\,,$$ 
\begin{equation}
\Dl\left(\vk-i\vep_r-\frac{i}{2}\vep_y\,,\,\vec y\right)
\bigg|_{\vec y=\r}\,\approx\,\Dl(\vk,\r)
\,=\,\dl(\hk\cdot\widehat\Dl_1(\r)+i\hk\cdot\widehat\Dl_2(\r))\,,\end{equation}
where $\dl$ is the gap amplitude, $\hk$ is unit reciprocal vector and the
orbital momentum vector is given by $\widehat\Dl_1\times \widehat\Dl_2=\hl$. 
The Eqs.(5) read that (2) can be written in the mixed representation in the 
form:
\begin{equation}
i\om\, g-\left(\matrix{ \xi-i c_{_F}\hk\cdot\vep &\Dl(\vk,\r)\cr
 \Dl^*(\vk,\r) &-\xi+ic_{_F}\hk\cdot\vep\cr}\right)g\,=\,\UN\,,
\end{equation}
where $g\equiv g_{_{\vk}}(\r),$ $\,\xi\approx c_{_F}(k-k_{_F})$, $c_{_F}$ is 
the Fermi velocity, and $\omega$ is fermionic Matsubara frequency. 

Thus we have obtained the approximate Eq.(6) which can be nicely treated as 
1-dimensional because the spatial differentiation is along the directions 
labeled by $\hk$. Indeed, in [26] a gradient expansion method is presented 
to study dynamics of spatially inhomogeneous systems provided inhomogeneities 
are slow compared to the relevant length scales. As the result, a separation 
of 3-dimensional problem into a collection of 1-dimensional subsystems occurs. 
Proofs useful for justification of our approach can be picked up from [26]. 

As far as we are interested in $\vj$ in arbitrary point, say, ${\cal O}$, let 
us define the spherical coordinates $\rho,\theta,\ph$ centered at it 
and linearize the slowly varying order parameter as follows:
\begin{equation}
\Dl(\vk,\r)\,\approx\,\Dl(\hk,\rho=0)+\al\rho\,\equiv\,\al(\rho_0+\rho)+i\Dl\,,
\end{equation}
$$
\Dl\,\equiv\,\IM\,\Dl(\hk,\rho=0)\,.
$$
As the physical result in ${\cal O}$ is assumed to be independent on the 
choice of the point, it can be calculated at any $\r$ with $\r\to{\cal O}$ in 
final formulas. Therefore, we shall solve (6) at $\r=\rho\hk$ so that 
$\hk\cdot\vec\cd$ is simply $\cd/\cd\rho$ and put $\rho=0$ in the result [10]. 

To be precise, we shall consider our problem for the coherence length 
$\xi_0$ much smaller than a length of the orbital vector $\hl$ variation:
$$
\xi_0\,=\,\frac{c_F}{\dl}\,\ll\,|\vep\otimes\hl|^{-1}\,,
$$ 
or
\begin{equation}
\frac{1}{\chi^2}\,\equiv\,\xi_0|\vep\otimes \hl |\,\ll\, 1
\end{equation}
(the London limit).
The parameter $\al$ (7) depends on the angle variables, vector's $\vec v_s$ 
components and first derivatives of $\hl$ taken in ${\cal O}$, and it will
be written explicitly in the Sec.~5. From (8) it is seen that the condition 
$\al\rho\,\lll\dl$ ensuring the linearization (7) implies $\rho/\xi_0\,\lll\,
\chi^2,$ and holds better provided $\chi^2$ is greater.

\def \x{{\text{x}}}
Changing the variable $\x=(\al/c_{_F})^{1/2}(\rho+\rho_0)$ and eliminating 
$\xi$ from the L.H.S. in (6) one gets:
\begin{equation}
(i\om+{\cal H})\,G\,=\,e^{i\text{x}\xi(\al c_{_{F}})^{-1/2}}\,\UN\,,
\end{equation}
where
\begin{equation}
{\cal H}\,=\,i\sqrt{\al c_{_F}}\si_3 \df -\sqrt{\al c_{_F}}\si_1 \x+\Dl\si_2
\end{equation}
is the Hamiltonian and $\si_i$ are the Pauli matrices. In this case (3) becomes
\begin{equation}
\vec j=k^3_{_F}(8\pi^3 c_{_F})^{-1}\int{\text d}\Om\, \hk\left(\be^{-1}
\sum\limits_\om{\cal J}\right),
\end{equation}
where ${\cal J}$ is the $\xi$-integrated Green function:
\begin{equation}
{\cal J}({\text x})=\int{\text d}\xi e^{-i \x\xi(\al c_{_F})^{-1/2}}\,
G_{11}(\x),
\end{equation}
and $G(\x)$ is to be determined from (9).

\section{SOLUTION OF THE NON--HOMOGENEOUS\- EQ\-UA\-TION IN FO\-RM OF SERIES}

To solve (9) let us take $G({\text x})$ in the form
\begin{equation}
G=\sqrt2\, {\text u}\left(\matrix {h_1&h_2\cr
f_1&f_2\cr}\right),\quad {\text u}\equiv\frac1{\sqrt2}\left(
\matrix {1&\,\,\,1\cr
i&-i\cr}\right),
\end{equation}
where $h_{1,\,2}\equiv h_{1,\,2}(\x)$ and $f_{1,\,2}\equiv f_{1,\,2}(\x)$ are 
now to be determined. Adjoint action ${\text u}^{-1}\si_1{\text u}\,=\,\si_2$ 
(cycl.perm.) of the unitary matrix ${\text u}$ on the Pauli matrices 
transforms ${\cal H}$ (10) to ${\cal H}_{em}$:
\begin{equation}
{\text u}^{-1}\,{\cal H}\,{\text u}\,=\,{\cal H}_{em},\quad
{\cal H}_{em}\,=\,\left(\matrix{\Dl& i\sqrt{\al c_{_F}}a^{-}\cr
-i\sqrt{\al c_{_F}}a^+  &-\Dl\cr}\right),
\end{equation}
where $a^{\pm}=\x\mp {\text d}/{\text dx}$. The operator 
${\cal H}_{em}$ reminds the Hamiltonian of spinning electron in constant 
homogeneous magnetic field. It is straightforward to obtain its eigen-values 
$E_0,\,\pm E_n$ and eigen-functions $\widehat\Psi_0,\,\widehat\Psi^{\pm}_n\,
(n\ge1)$ [22]:
$$
\widehat\Psi_0\,=\,\left(\matrix {0\cr
\psi_0(\x)\cr}\right),\quad E_0=-\Dl,$$
\begin{equation}
\widehat\Psi^{(s)}_n\,=\,\frac1{\sqrt {2 E_n}}
\left(\matrix {\,\,\,\,\,\sqrt{E_n+s\Dl\,}\,\,\psi_{n-1}(\x) \cr
-is\sqrt{E_n-s\Dl\,}\,\,\psi_n(\x)\cr}\right),\quad sE_n,
\end{equation}
where $s=\pm,\,\,E_n=\sqrt{\Dl^2+2\al c_{_F}n\,}$ and $\psi_n(\x)$ are the 
Hermite functions.

Let us use (14) to pass from (9) to the equation
\begin{equation}
(i\om+{\cal H}_{em})\left(\matrix{ h\cr
f\cr}\right)\,=\,\dl({\text x-x'})\,e^{i\x\xi(\al c_{_F})^{-1/2}}
\left(\matrix{1/2\cr 1/2\cr}\right)\,,
\end{equation}
to calculate $G_{11}$. The Dirac $\dl$-function is placed in the R.H.S. of 
(16), the unknown $h,f$ depend now on ${\text x, x}'\,$ and the required 
entry is given by 
\begin{equation}
G_{11}({\text x})=\int{\text dx}'(h({\text x,x}')
\,+\,f({\text x,x}'))\,.
\end{equation}
To solve (16) it is natural to expand $\left(\matrix{ h\cr f\cr}\right)$ in 
the functions (15):
\begin{equation}
\left(\matrix{ h\cr
f\cr}\right)({\text x,x}')\,=\,B({\text x}')\,\widehat\Psi_0({\text x})\,+\,
\sum\limits_{s=\pm}\,\sum\limits^\infty_{n=1}
b^{(s)}_n({\text x}')\,\widehat\Psi^{(s)}_n({\text x})\,.
\end{equation}
We calculate $B({\text x}')$, $b^{(s)}_n({\text x}')$ using orthogonality of 
the vectors (15) [22,~24], and from (12),~(17),~(18) obtain $\cal J$:
\begin{equation}
{\cal J}\,=\,\pi\sqrt{\al c_{_F}}\left[\frac{<\widehat\Psi_0,\widehat\Psi_0>}
{i\om+E_0}\,+\,\sum\limits_{s=\pm}\,
\sum\limits^\infty_{n=1}\frac{<\widehat\Psi^{(s)}_n,\widehat\Psi^{(s)}_n>}
{i\om+sE_n}\right],
\end{equation}
where $<\cdot,\cdot>$ stands for Hermitian scalar product. The representation 
(19) for the $\xi$-integrated Green function is alternative to that which has 
been found in [10] (the Eq.(34)) as a quadratic combination of parabolic 
cylinder functions.

Now summation over $\om$ is straightforward [27] and one gets:
$$
\be^{-1}\sum\limits_\om{\cal J}\,
=\,\pi\sqrt{\al c_{_F}}
\left[n(E_0)\psi^2_0\,+\,\frac12\sum\limits^\infty_{n=1}
\left(\psi^2_{n-1}+\psi^2_n\right)\,+\right.$$
\begin{equation}
\left.+\,\,\frac\Dl2\sum\limits^\infty_{n=1}\left(\psi^2_{n-1}-\psi^2_n\right)
\frac{\tanh(\be E_n/2)}{E_n}\right]\,,
\end{equation}
where $n(\ep)$ is the Fermi weight. Inserting (20) to $\vj$ (11) we obtain the 
required general representation for the mass current near zero temperature 
[23,~24]. Due to the explicit dependence on $\be$, the Eq.(20) admits $T\to 0$ 
as well: one has to replace $n(E_0)$ by the Heavyside function $\theta(E_0)$ 
and $\tanh(\be E_n/2)$ by 1. It can be argued that the second term in (20) is 
irrelevant with regard to the angle integration and therefore $\vj$ acquires 
the final form as the series: 
\begin{equation}
\vj=k^3_{_F}(8\pi^2 c_{_F})^{-1}\int{\text d}\Om\,\widehat k
\sqrt{\al c_{_F}}\,\bigg[n(E_0)\psi^2_0
\,+\,\frac\Dl2\sum\limits^\infty_{n=1}\left(\psi^2_{n-1}-\psi^2_n\right)
\frac{\tanh(\be E_n/2)}{E_n}\bigg].
\end{equation}

\section{INTEGRAL REPRESENTATIONS AND THE\-IR LIMITING CASES}

\subsection{INTEGRAL REPRESENTATIONS}
\medskip

Practically, it is more convenient to be concerned with an integral 
representation equivalent to (21). Such representation has been found in [23] 
at zero temperature, and here we shall deduce it for $\be^{-1}\sum\limits_\om
{\cal J}$ and $\vj$ at $T\ne 0$ [24]. These representations will admit special 
limiting cases. 

First of all, one has to rearrange (19) [22]:
\begin{equation}
{\cal J}\,=\,-\pi\sqrt{\al c_{_F}}\,\sum\limits^\infty_{n=0}\psi^2_n
\left(\frac{i\om-\Dl}{\om^2+E^2_{n+1}}\,+\,
\frac{i\om+\Dl}{\om^2+E^2_n}\right).
\end{equation}
By the formula 
$$
\be^{-1}\sum\limits_\om i\om(\om^2+\ep^2)^{-1}=-1/2\,
$$
we see that the odd in $\om$ part of (22) is responsible for the second term 
in (20) and it is enough to consider only the ``even'' part of (22):
\begin{equation}
{\cal J}_e\,=\,\frac\pi2\,\frac\Dl{\sqrt{\al c_{_F}}}\,\sum\limits^\infty 
_{n=0}\psi^2_n\,\left[(|\la|^2+n+1)^{-1}-(|\la|^2+n)^{-1}\right],
\end{equation}
where $|\la|^2\equiv(\om^2+\Dl^2)(2\al c_{_F})^{-1}$.

By the formula (AI.2) (${\sl APPENDIX~I}$) the series (23) can be expressed 
as the integral
\begin{equation}
{\cal J}_e\,=\,-\Dl\left(\frac\pi{\al c_{_F}}\right)^{1/2}
\int\limits^\infty_0 {\text d}t\,(\tanh t)^{1/2}\,e^{-\text{x}^2\tanh t-2|\la|^2\,t}\,.
\end{equation}
Thus one can go further:
$$
\be^{-1}\sum\limits_\om {\cal J}_e\,=\,-\Dl\left(\frac\pi{\al c_{_F}}
\right)^{1/2} \int\limits^\infty_0 {\text d}t\,(\tanh t)^{1/2}
$$
\begin{equation}\times\big(T\,\vartheta_2(0,\,i\tau)\big)
\,e^{-{\text x}^2\tanh t-(\Dl^2/\al c_{_F})\,t}\,,
\end{equation}
where the elliptic theta function $\vartheta_2$ [28] implies the series
$$
\sum\limits^\infty_{m=0}a^{(m+\frac12)^2}=\frac12\vartheta_2(0,i\tau)\,,
$$
$\tau=(-1/\pi)\log a$, and $a=\exp(-4\pi^2 T^2 t/\al c_{_F})$. Changing the 
integration variable $t\longmapsto \kappa\,t$, $\kappa\,=\,\al 
c_{_F}(\be/2)^2$, one can rewrite (25) more suitably for studying the 
limiting cases below:
$$
\be^{-1}\sum\limits_\om {\cal J}_e\,=\,-\Dl\,\frac{\,\kappa^{1/2}\,}2 
\int\limits^\infty_0 {\text d}t\left(\frac{\tanh (\kappa t)}{t}\right)^{1/2}$$
\begin{equation}
\times\widetilde{\Theta}(t)
\,e^{-{\text x}^2\tanh (\kappa t)-(\Dl\be/2)^2\,t}\,,
\end{equation}
where $\,\widetilde{\Theta}(t)\,=(\pi t)^{1/2}\,\vartheta_2(0,\,i\,\pi \,t)$.

The Eqs.~(25) and (26) are just to be substituted to (11) to get the general 
integral representations for $\vj$ near zero temperature. These representations 
are very convenient in calculating higher corrections to (1). Before to proceed to 
it in the Sec.5, let us consider some particular limits for (25), (26).

\subsection{LIMITING CASES}

Let us represent the situation by the following ``commutative diagram'':
$$
\vj=\vj\big|_{\be^{-1}\sum{\cal J}=(26)}\qquad\stackrel{1}{\longrightarrow}\qquad {\text I} \\
$$
\begin{equation}
{\text 4}\bigg\downarrow\qquad\qquad\qquad\qquad\qquad\qquad\bigg\downarrow{\text 2}\\
\end{equation}
$$
\qquad{\text II}\qquad\qquad\qquad\qquad\stackrel{3}{\longrightarrow}\qquad{\text (1)}
\quad\quad, 
$$
where the horizontal arrows $1, 3$ mean $T\to 0$ and the vertical ones 
$2, 4$ -- the lowest ``gradient'' approximation.

To begin with, the usage of the limit 
$$\lim\limits_{\tau\to0}\,\vartheta_2(0,i\tau)\,=\,\tau^{-1/2}\,$$
[28] in (25) allows $\vj$ to be written at $T=0$ as follows (the arrow 
${\sl 1}$):
\begin{equation}
\vj\,=\,-\,3\rho\,(8\pi c_{_F})^{-1}\,
\int{\text d}\Om\,\hk\,\Dl\,F\,(\x^2,\,\Dl^2/\al\,c_{_F})\,,
\end{equation}
where $\rho\,=\,k^3_{_F}/3\pi^2$ (two spin projections are taken into 
account), and the function $F(p,q)$ is given by
$$
F(p,\,q)\,=\,\int\limits^\infty_0
{\text d}t\,\left(\frac{\tanh t}t\right)^{1/2} \,e^{{-p\tanh 
t}-q\,t}\,,\quad\quad q\,>\,0\,.
$$

To perform the lowest ``gradient'' approximation (the arrow ${\sl 2}$), we 
replace tanh $t$ by $t$ in $F(p,\,q)$ (see (8) and (AII.6)) and (28) takes the
form:
\def\tanhs{\text{\rm tanh}}
\def\sh{\text{\rm sh}}
$$
\vj\,=\,-3\rho\,(8\pi c_{_F})^{-1} \int{\text d}\Omega\,\hk \Delta
\int\limits^\infty_0 {\text d}t\,e^{-\left(\x^2+\frac{\Dl^2}
{\alpha c_{_F}}\right)\,t}\,=
$$
\begin{equation}
=\,-\,3\rho\,(8\pi)^{-1} \int{\text d}\Omega\,\hk\, 
          \frac{\alpha \Dl}{\al c_{_F} \x^2+\Dl^2}
\end{equation}
(compare with the Eq.(44) in [10]). To understand (29), let us recall the 
representation for $\vj_0$ which has been discussed in [19,~20]:
\begin{equation}
\vj_0\,=\,3\rho\,(8\pi)^{-1}\int{\text d}\Omega\,\hk
\left((\hk \cdot\vec\cd) \arctan\left(\frac{\hk\cdot \widehat\Dl_2}{\hk
\cdot\widehat\Dl_1}\right)\right)\,,
\end{equation}
and which results in (1) with $\,\hl= \widehat\Dl_1\times\widehat\Dl_2$ and 
$(v_s)_i =2^{-1}\,\widehat\Dl_1\cdot\partial_i\widehat\Dl_2$. Using 
$$
\hk\cdot\vec\cd\,=\,\sqrt{\frac{\al}{c_{_F}}}\,\df\,,\quad\quad
\frac{\,\hk\cdot\widehat\Dl_2}{\,\hk\cdot\widehat\Dl_1}\,=
\,\frac{\Dl}{\x\sqrt{\al c_{_F}}}\,\,,
$$
it is easy to check coincidence of (29) and (30). So without any special gauge 
for $\al$ it is seen that (1) is the lowest London limit approximation to (28).

To consider the steps ${\sl 4} \to {\sl 3},$ one should replace $\tanh 
(\kappa t)$ by $\kappa t$ in (26) (the arrow ${\sl 4}$) due to steepest 
descent validity at $\Dl\ne 0$ for large $\be:$ 
\begin{equation}
\be^{-1}\sum\limits_\om {\cal J}_e\,=\,-\,\frac{\al c_{_F}\Dl}{8}\,\be^2 
\int\limits^\infty_0 {\text d}t\,\widetilde{\Theta}(t)\, e^{-(|\Dl|\be/2)^2\,t}
\,.\end{equation}
The R.H.S. of (31) is the Laplace transform of $\widetilde{\Theta}(t),$ 
and can be expressed through the so-called Yosida function  $\text{Y}$ [29]:
\begin{equation}
a^2\int\limits^\infty_0 {\text d}t\, \widetilde{\Theta}(t)\,e^{-a^2t}\,=\,
1-\int\limits^\infty_0\frac{dy}{\cosh^2\sqrt{y^2+a^2}}\equiv\,1-\text{Y}(a).
\end{equation}
To check (32) it is enough to integrate its L.H.S. as the series and 
to re-express the answer by the Poisson summation formula, whereas 
$\cosh^{-2}(\text{y})$ in the R.H.S. has to be expanded first 
in $\exp\,(-2\text{y})$ and then integrated [24]. Using (32) one obtains:
\begin{equation}
\vj\,=\,- 3\rho\,(8\pi)^{-1}\int{\text d}\Om\, \hk \,\frac{\al \Dl}{|\Dl|^2}
\left(1-\text{Y}\left(\frac{|\Dl|\be}{2}\right)\right)
\end{equation}
(the point $II$ on (27)). The Eq.(33) is just the leading $\vj_0$ ``dressed'' 
by thermal corrections which has been found by Cross [8]. At $T=0$ 
(the arrow $3)$ $\text{Y}(\infty)=0$ and we recover (1).

\section{EXPLICIT CALCULATIONS}

This section is devoted to the main problem of the present paper. That is it
will be concerned with asymptotical expansion of $\vj$ (28) in order to deduce 
the London limit corrections to (1). At fixed $\hk$ the overall phase of the
order parameter $\Dl(\hk,\,\r)$ can always be changed to make $\al$ (7) a real 
positive. Thus the Eq.(7) can be thought of as
\begin{equation}
\exp(-i\psi)\,\Dl(\hk,\,\r)\,\equiv\,\Dl_0\,+\,\al\,\rho\,,
\end{equation}
where
\begin{equation}
\al\,=\,\dl\,{\cal{M}}\,\exp (i(\pi/2-\psi))\,.
\end{equation}
$$
\Dl_0\,=\,\dl\,\sin\theta\,\exp (i(\phi-\psi))\,.
$$
In (34), (35) the phase $\psi$ is to be adjusted while ${\cal M}$ is given by
$$
{\cal{M}}\,=\, -\cos^2\theta\,\cd_3\wh l_2\,+\,2m\,\sin^2\theta
\,e^{i\phi}\,(v_1\cos \phi +v_2\sin \phi)\,+$$
$$+\,\sin\theta\cos\theta\, e^{i\phi}\,(2mv_3+\frac{\,i\,}2\,{\text div}\,\hl
-\frac{\,1\,}2\,\hl\cdot\rot\,\hl)\,+$$
\begin{equation}
+\,\frac{\,1\,}2\,\sin \theta\cos\theta\,e^{-i\phi}\,(-\cd_1\wh l_2-\cd_2
\wh l_1\,+\,i(\cd_1\wh l_1-\cd_2\wh l_2))\,,
\end{equation}
where $\vec v\equiv\vec v_s$ ({\sl APPENDIX II}). Without loss of generality 
$\widehat\Dl_2({\cal O})$ can be chosen along $\hl\times\rot\,\hl$ so that 
$\cd_3\wh l_1\,,\,\cd_3\wh l_3$ become zero and thus ${\text div}\,\hl=
\cd_1\wh l_1 + \cd_2\wh l_2$ [10]. Besides, $\hl\cdot\rot\,\hl$ is $\cd_1
\wh l_2-\cd_2\wh l_1$ once the third axis is chosen along $\hl({\cal O})$. 
Moreover, $\cd_1\wh l_1,\cd_2\wh l_2$ can be excluded from the consideration 
[10]. Therefore, apart of $v_1$ and $v_2$, only $\cd_1\wh l_2 + \cd_2\wh l_1$, 
$\,2mv_3-$ $(1/2)\,
\hl\cdot\rot\,\hl$, and $\cd_3\wh l_2=\rot\,\hl\times\hl$ are the relevant 
gradient combinations. Besides, no difference is expected once ${\cal M}$
is considered as dependent separately either on $\cd_1\wh l_2+\cd_2\wh l_1$ or 
$\,2mv_3-$ $(1/2)\,\hl\cdot\rot\,\hl$.

To make calculations manageable it is appropriate to put a part of gradients 
in ${\cal M}$ equal to zero so to consider the dependence of $\vj$ on the 
remaining ones. Clearly, it is not necessary to enumerate all the possible 
cases, but it is enough to point out rather characteristic combinations. To 
this end, let us take ${\cal M}$ in the following reduced form:
$$
{\cal{M}}\,=\,-\,\cd_3\wh l_2\,\cos^2\theta\,+\,
(2mv-\cd_1\wh l_2)\,\sin\theta\cos\theta\,e^{i\phi}$$
\begin{equation}
\equiv\,-\,\frac1{\xi_0\chi_1^2}\cos^2\theta\,+\,\frac1{\xi_0\chi^2_2}
\sin\theta\,\cos\theta\,e^{i\phi}
\end{equation}
(at $\cd_1\wh l_2+\cd_2\wh l_1=0,$ $v\equiv v_3$). Once $\psi$ is obtained 
explicitly so that $\al\in\BR^+$ we get:
$$
\al c_{_F}\,=\,\frac{\,\dl^2\,\sin^2\theta\,}{Q}\,,\quad
x_0^2\,=\,\left(\frac{\sin\phi}{\chi_1^2\,\tan^2\theta}\right)^2\,\,Q^3\,,
$$
\begin{equation}
\dl^{-1}\,\frac\Dl{\,Q\,}\,=\,\cos\theta\,\left(\frac{\,1}{\chi^2_1}\,\frac
{\,\cos\phi\,}{\tan\theta}\,-\,\frac{\,1\,}{\chi^2_2}\right)\,,\end{equation}
where
$$
\frac{\,1\,}{Q^2}\,=\,\left(\frac1{\chi_2^2}\,\frac{\sin\phi}{\tan\theta}
\right)^2\,+\,\left(\frac1{\chi_2^2}\,\frac{\cos\phi}{\tan\theta}\,-\,
\frac1{\chi_1^2}\,\frac1{\tan^2\theta}\right)^2\,.
$$

A convenience is apparent after [24] to integrate by parts in $F(p,q)$ in (28)
so that
\begin{equation}
F(\x^2,\,\Dl^2/\al c_{_F})\,=\,\frac{\,1\,}Q\,\left(1\,+\,\Phi({\x}^2,\,Q)
\right)\end{equation}
with
\begin{equation}
\Phi({\x}^2,\,Q)\,=\,\int\limits^\infty_0\,e^{-\,Q\,t}
\left(\sqrt{\frac{\tanh t}t} \,e^{{\x}^2(t-\tanh t)}\right)^\prime{\text d}t\,.
\end{equation}
We have used in (39), (40) the following circumstance. According to the 
Eq.(34), the order parameter in ${\cal O}$ is $\Dl_0$ and the square of its 
modulus has the simple form:
$$|\Dl_0|^2\,\equiv\,|\Dl|^2\,=\,\Dl^2\,+\,\al c_{_F}
{\x}^2\,=\,\dl^2\sin^2\theta$$
by the Eqs.(7) and (35). Now, from (38) it is seen that $Q$ in (39),
(40) is just $|\Dl|^2/\al c_{_F}$. The first term in (39) is responsible for 
$\vj_0$ (see (29)), while the second one -- for the total correcting 
contribution:
\begin{equation}
\vj_{\text{corr}}\,=\,-\,3\rho\,(8\pi)^{-1}\,\int{\text d}\Om\,\hk\,
\frac{\al\Dl}{|\Dl|^2}\,\Phi\,({\text x}^2,\,Q)\,.
\end{equation}

In what follows we shall investigate (41) with ${\cal M}$ (37):
\begin{equation}
\vj_{\text{corr}}\,=\,3\rho\,(8\pi\xi_0)^{-1}\,\int{\text d}\Om\,\hk\,\cos\theta
\,\left(\frac{\,1\,}{\chi_2^2}\,-\,\frac{\,1\,}{\chi^2_1}\frac{\cos\phi}
{\tan\theta}\right)\,\Phi\,\left({\x}^2\,=\,{\x}^2_0\,,\,Q\right)\,.
\end{equation}
Varying $\chi_1$, $\,\chi_2$ in (42) the following three cases ({\sl Examples} 
{\sl 1,\,2},\, and {\sl 3}) can be obtained. Fixing $\chi_2$ (or 
$\chi_1$) $\,\gg 1$ and tending $\chi_1$ (or $\chi_2$) to infinity we shall 
get {\sl Example 1} (or {\sl Example 2}). Taking $\chi_1=\chi_2=\chi
\gg 1$ we shall arrive to {\sl Example 3}. Each time our attention will be 
called to the quadratic and cubic contributions to $\vj_{\text{corr}}$, i.e. 
to the terms proportional to $(\xi_0\chi^4)^{-1}$ and $(\xi_0\chi^6)^{-1}$. 
Clearly, the case {\sl 1} corresponds to $\rot\,\hl$ parallel 
to $\hl\,$ and the case {\sl 2} --- to $\rot\,\hl$ perpendicular to $\hl$. 
Therefore, the case {\sl 1} implies all the three contributions in (1), while 
the second one corresponds only to the pure orbital content of (1).

\medskip
\subsection{EXAMPLE 1:~rot$\,\hl$ IS PARALLEL TO $\hl$.}

Here only $2mv-\cd_1\wh l_2\ne0$ in ${\cal M}$ (37). As far as we deduce from (38)
that $$
Q\,=\,\chi^2\,|\tan\theta|\,,\quad \x_0\,=\,0\,,$$
where $(\xi_0\chi^2)^{-1}\equiv 2mv-\cd_1\wh l_2\,>\,0$, the Eq.(42) reads only 
the third component, say, $j$ to be nonzero now:
\begin{equation}
j\,=\,\frac{3\rho}{2}\,\frac1{\xi_0\chi^2}\int
\limits^\infty_0\,{\cal F}(\chi^2u)\,
\frac{u\,{\text d}u}{(u^2+1)^{5/2}}\,,\end{equation}
where
$${\cal F}(\chi^2u)\,=\,\int\limits^\infty_0\,e^{-\chi^2ut}\left(
\sqrt{\frac{\tanh t}{t}}\right)^\prime\text{d}t\,,
$$
and $u=|\tan\theta|$. Concrete form of ${\cal F}$ is not of importance for 
us. It is enough to know that ${\cal F}(s)\to\,$const as $s\to 0$, and 
\begin{equation}
{\cal F}(\chi^2u)\,=\,\frac{{\text a}}{(\chi^2u)^2}\,+\,
                     \frac{{\text b}}{(\chi^2u)^4}\,+\,...\,,\quad \chi^2u\gg 1\,,
\end{equation}
where a$=-1/3$.

To do the estimation let us break the integral over $u$ into two parts:
\begin{equation}\int\limits^\infty_0\,{\cal F}(\chi^2u)\,
\frac{u\,{\text d}u}{(u^2+1)^{5/2}}\,=\,U_1\,+\,U_2\,,
\end{equation}
where
$$
U_2\,=\,\int\limits^\infty_1\,{\cal F}(\chi^2u)\,\frac{u\,{\text d}u}
{(u^2+1)^{5/2}}\approx\,\frac{\text a}{\chi^4}
\,\int\limits^\infty_1\,(u^2+1)^{-5/2}\,\frac{\,{\text d}u\,}{u}\,,
$$
because $\chi^2u\gg 1$ is valid, and we are interested in
contributions of total degree in $\chi$ not less than $-6$. As to $U_1$,
$$
U_1\,=\,\frac1{\chi^4}\int\limits^{\chi^2}_0\,{\cal F}(u)\,
\frac{\,u\,{\text d}u\,}{(1+u^2/\chi^4)^{5/2}}\,,
$$
its denominator can be expanded in powers of $u^2/\chi^4$ so that $U_1$ will 
acquire the form of series where each term is given by the appropriate 
integral. The upper bound $\chi^2$ of all these integrals can be extended to 
infinity provided the integral is convergent, either some regularization by 
counter-term is needed. It is easy to see from (44) that $\int {\cal F}u
\,{\text d}u$ means only the logarithmic divergency, $\int {\cal F}u^3{\text d}
u$ --- both the quadratic and logarithmic ones and so on.

Let us do the first subtraction writing 
\def \X{{\text X}}
\def \Y{{\text Y}}
\def \Z{{\text Z}}
$U_1\,=\,\X\,+\,\Y\,$, where
\begin{equation}
\X\,=\,\frac1{\chi^4}\int\limits^{\chi^2}_0\,{\cal F}(u)
\left(\left(1+\frac{u^2}{\chi^4}\right)^{-5/2}-1\right)\,u\,{\text d}u\,,
\quad\Y\,=\,\frac1{\chi^4}\int\limits^{\chi^2}_0\,{\cal F}(u)\,u\,{\text d}u\,.
\end{equation}
Once the integral $\Y$ is divergent logarithmically at $\chi^2\to\infty$, it 
can be represented approximately:
$$
\Y\,\approx\,\frac a{\chi^4}\,\log\chi^2\,+\,\frac1{\chi^4}
\int\limits^{\infty}_0\left(u\,{\cal F}(u)-
\frac{{\text a}}{u+1}\right){\text d}u\,.
$$

Let us turn to $\X$ (46). The asymptotic (44) tells us that $\X$ is divergent
at $\chi\to\infty$, and the whole (44) is needed for regularization. It is not
difficult to realize that the total contribution of the order $\chi^{-4}$
appears as that counter-term which results from $\X$ once ${\cal F}$ 
\def \a{\text a}
\def \b{\text b}
is replaced by ${\a}/u^2$. Therefore,
$$
U_2\,+\,\X\,=\,\frac{{\text a}}{\chi^4}\left(-\frac43\,+\,\log\,2\right)\,.$$
Therefore at $\a=-1/3$
\begin{equation}
U_1\,+\,U_2\,=\,\frac1{3\chi^4}\left(\frac43\,+\,
\int\limits^{\infty}_0\left(3u\,{\cal F}(u)\,+\,\frac1{u+1}\right){\text d}u
\,+\,\log\frac{\,1\,}{2\chi^2}\right)\,.
\end{equation}
Finally, the use of the Eqs.(43), (45) and (47) enables the third component of 
$\vj$ to be completly written as follows:
\begin{equation}
j_3\,=\,\frac{\,\rho\,}2\,\frac1{\xi_0\chi^2}\,\left(1\,+\,
\frac1{\chi^4}\,\log\frac{\,{\cal B}\,}{\chi^2}\right)\,,
\end{equation}
where
$$
\log 2{\cal B}\,=\,\frac43\,+\,\int\limits^{\infty}_0\left(3u\,{\cal F}(u)
\,+\,\frac1{u+1}\right){\text d}u\,.
$$
All the quadratic corrections predicted in [10] are zero in the present gauge, 
and the lowest one turns out to be cubic with the logarithm of the London 
parameter. 

\def \tanhs{\text{\rm tanh}}
\def \sh{\text{\rm sh}}

\subsection {EXAMPLE 2:~rot$\,\hl$ IS PERPENDICULAR TO $\hl$}

In this case the Eqs.(38) result in
$$
Q\,=\,(\chi \tan\theta)^{2}\,,\quad\quad
\x_0\,=\,-\chi|\tan\theta|\sin\phi\,,$$
where $(\xi_0\chi^2)^{-1}\,\equiv\,\cd_3\wh l_2\,>\,0\,$. From (42) the components 
2 and 3 of $\vj_{\text{corr}}$ are zero, whereas the first one acquires the 
form:
\begin{equation}
j\,=\,-\,\frac{\,3\rho\,}4\,\frac{\,1\,}{\xi_0\chi^2}\int\limits^\infty_0
\,{\cal F}(\chi^2u^2)\,\frac{u\,{\text d}u}{(u^2+1)^{5/2}}\,,
\end{equation}
where
$$
{\cal F}(\chi^2u^2)\,=\,\int\limits^\infty_0\,e^{-\chi^2u^2t}
\left(\sqrt{\frac{\tanh t}{t}}\,_1F_1(\,\frac{\,1\,}2;\,2;\,\chi^2u^2(t-
\tanh t))\right)^\prime{\text d}t\,,
$$
$u=|\tan\theta|$ and the formula 
\begin{equation}
\int\limits^1_0 {\text d}v\,\sqrt{1-v^2}\,e^{v^2p}\,=\,\frac\pi4\,_1F_1
\left(\,\frac{\,1\,}2,\,2;\,p\right),
\end{equation}
$$
p\,=\,u^2(t-\text{\rm tanh\, t}\,)\,\ge\,0\,
$$
is used to re-express the integration over $v=\sin\phi$. In (50) $_1F_1$ is 
the Kummer function [28]. The relevant analytical properties of ${\cal F}$ 
are the following: ${\cal F}(0)=\,$const and asymptotically
$$
{\cal F}(\chi^2u^2)\,=\,\frac{\,{\a}\,}{(\chi u)^4}\,+\,
\frac{\,{\b}\,}{(\chi u)^8}\,+...\,,\qquad\,\chi^2u^2\gg 1\,,
$$
where $\a=1/6$.

Again let us represent the integral 
$$
\int\limits^\infty_0\,{\cal F}(\chi^2u^2)\,
\frac{u\,{\text d}u}{(u^2+1)^{5/2}}\,,
$$
as the sum of $U_1$ and $U_2$ so that
$$
U_2\,=\,\int\limits^\infty_1\,\,{\cal F}(\chi^2u^2)\,
\frac{u\,{\text d}u}{(u^2+1)^{5/2}}\,\approx\,\frac{\,\a\,}{\chi^4}\,
\int\limits^\infty_1\,(u^2+1)^{-5/2}\,\frac{\,{\text d}u\,}{u^3}\,.
$$
Now two first subtractions are needed to estimate $U_1\,=\,\X\,+\,\Y\,+\Z\,$, 
where
$$
\Z\,=\,\frac{\,1\,}{\chi^2}\int\limits^{\chi}_0\,{\cal F}(u^2)\,u{\text d}u\,,
\qquad
\Y\,=\,-\,\frac{\,5\,}2\,\frac{\,1\,}{\chi^4}\int\limits^{\chi}_0\,
{\cal F}(u^2)\,u^3{\text d}u\,,
$$
$$
\X\,=\,\frac{\,1\,}{\chi^2}\int\limits^{\chi}_0\,{\cal F}(u^2)
\left(\left(1+\frac{u^2}{\chi^2}\right)^{-5/2}\!\!\!-\,1\,+\,
\frac{\,5\,}2\frac{\,u^2\,}{\chi^2}\right)u\,{\text d}u\,.
$$

Clearly, $\Z$ is convergent at large $\chi$ and approximately
$$
\Z\,\approx\,\frac1{\chi^2}\int\limits^{\infty}_0\,{\cal F}(u^2)\,u\,
{\text d}u\,-\,\frac{\,\a\,}{2\,\chi^4}\,.$$
Further, a single counter-term is required for $\Y$:
$$\Y\,\approx\,-\,\frac{\,5\,}2\frac{\,1\,}{\chi^4}\,\int\limits^{\infty}_0
\left(u^3\,{\cal F}(u^2)\,-\,\frac{\,a\,}{u+1}\right){\text d}u\,-\,
\frac{\,5\,\a\,}2\,\frac{\,\log\chi\,}{\chi^4}\,.$$

Now let us consider $\X$. Here the series in the brackets begins with the term 
proportional to $(u/\chi)^4$, and a single regulator is needed. The next term 
will require two and so on. The total contribution of the order $\chi^{-4}$ is 
given once ${\cal F}$ is replaced by $a/u^4$ in $\X$. The net result reads:
$$
U_2\,+\,\X\,=\,\frac{\,\a\,}{\chi^4}\left(\frac{\,37\,}{12}\,-\,
\frac{\,5\,}2\,\log 2\right)\,,
$$
and, therefore,
$$
U_1\,+\,U_2\,=\,\frac1{\chi^2}\,\int\limits^{\infty}_0\,{\cal F}(u^2)\,u\,
{\text d}u\,+\,\frac{\,\a\,}{\chi^4}\,\left(\frac{\,31\,}{12}\,+\,\frac{\,5\,}
2\,\log\frac1{2\chi}\right)\,-$$
$$-\,\frac{\,5\,}{2\chi^4}\,\int\limits^{\infty}_0\left(u^3{\cal F}(u^2)\,-\,
\frac{\,{\a}\,}{u+1}\right){\text d}u\,.
$$

As the final result, the non-zero part of $\vj$ is:
\begin{equation}
j_1\,=\,-\frac{\,\rho\,}4\,\frac1{\xi_0\chi^2}\,\left(1\,+\,\frac{\,{\cal A}\,}
{\chi^2}\,+\,\frac{\,5\,}{8}\,\frac1{\chi^4}\,\log\frac{\,{\cal B}}{\chi^2}
\right)\,,
\end{equation}
where
\begin{equation}
{\cal A}\,=\,3\,\int\limits^{\infty}_0\,{\cal F}(u^2)\,u\,{\text d}u\,\,
\approx\,-\,2\,\times\,10^{-1}\,,
\end{equation}
$$
\log 4{\cal B}\,=\,\frac{\,31\,}{15}\,-\,12\,\int\limits^{\infty}_0
\left(u^3{\cal F}(u^2)\,-\,\frac1{6(u+1)}\right){\text d}u\,.
$$
In this case there are two corrections, and the lowest is of the type 
$(\rot\,\hl)_\perp\lrot$ found in [10] for the current perpendicular to $\hl$ 
(one should be referred to the formula (53) in [10]). The coefficient 
${\cal A}$ (52) has been estimated numerically in [23]. The next term is the 
new cubic one and it includes the logarithm of the London parameter.

\subsection {EXAMPLE 3}

In this case we shall take into account the whole (37) which would imply 
appearance of the quadratic corrections of the type 
$\lrot(\vec v-(1/4m)\rot\hl)_{||}$ and $\lrot(\cd_1\hl_2+\cd_2\hl_1)$ [10]. 
However, we put here $\chi_1=\chi_2$ 
for simplicity, and so the answer expected would demonstrate such corrections
only in principle. To this end we shall investigate the third component 
$j_{\text{corr},3}$ which is along $\hl$. We obtain from (42):
\def \Q{{\cal Q}}
\begin{equation}
j\,=\,\frac{\,3\rho\,}{2\pi}\,\frac{\,1\,}{\xi_0\chi^2}
\int\!\!\!\!\int\limits_\prod\,\frac{{\text d}u{\text d}v}
{\,\sqrt{(u^2+1)^5\,(1-v^2)\,}}
\left((u\,-\,v){\cal F}_-\,+\,(u\,+\,v){\cal F}_+\right)\,,
\end{equation}
$$
{\cal F}_{\mp}\,=\,\int\limits^\infty_0\,\exp\left(-t\,(\chi u)^2\,\Q \right)$$
\begin{equation}
\times\,\left(\sqrt{\frac{\,\tanh t\,}t}\,\exp
\left((t\,-\,\tanh t)(\chi\,u)^2(1-v^2)\,\Q^3\right)\right)^\prime
{\text d}t\,,
\end{equation}
where $\Q^{-2}$ stands for
$$
\Q_{\mp}^{-2}\,=\,1\,+\,u^2\,\mp\,2\,u\,v\,=\,1\,-\,v^2\,+\,(u\mp\,v)^2\,,
$$
the domain $\prod$ is given by $\{(u,v):\,$ $u\in[0,\infty[,v\in[0,1]\}$,
and $u=\tan\theta$, $v=\cos\phi$.  

The function ${\cal M}$ is still rather complicated and so the present 
consideration becomes less elegant then the two previous. The estimations we 
are interested in will be obtained without providing the asymptotic integral 
formulas for the coefficients. Besides, we shall assume that not only $\chi$ 
but $\log\chi$ also is large (logarithmic accuracy), and thus only the 
logarithmic terms will be kept in the third order. Let us proceed estimating
$\Phi$ (40) in general situation. By steepest descent we get:
\begin{equation}
\Phi\,({\x}^2,Q)\,\simeq\,-\,\frac{\,1\,}3\,\frac1{\,Q^2\,}\,+\,2\,
\frac{\x^2}{\,Q^3}\,,
\end{equation}
at $Q=|\Dl|^2/\al c_{_F}\rrr 1$ (i.e. either $\Dl^2/\al c_{_F}$ or $\x^2$ 
must be $\rrr 1$, and it is forbidden to tend $\Dl^2/\al c_{_F}$ to zero). In 
the opposite case $Q < 1$
\begin{equation}
\Phi\,(\x^2,\,Q)\,\simeq\,-\,1\,+\,\pi^{1/2}\,{\x}^2\,\left(\frac
{\al c_{_F}}{\Dl^2}\right)^{1/2}\,.
\end{equation}

First of all, let us consider the contribution to $j$ (53) which is due to
$u\in [1,\infty[$. Here the function ${\cal F}$ can be expanded by steepest 
descent because $(\chi u)^2\Q\gg 1$. This expansion will begin with the 
third order term $const\times(\xi_0\chi^6)^{-1}$ which is not of 
interest for us. So, in what follows we shall take $0\le u\le 1$ in $j$.

Now let us consider the domain $0\le u\le 1/\chi$. Approximately we put:
\begin{equation}
\frac{\,3\rho\,}{2\pi}\,\frac{\,1\,}{\xi_0\chi^2}
\int\limits^{1/\chi}_0{\text d}u\int\limits^1_0\,\frac{{\text d}v}
{\,(1-v^2)^{1/2}\,}\,\left((u\,-\,v){\cal F}_-\,+\,(u\,+\,v){\cal F}_+
\right)\,.
\end{equation}
In this case $\Dl^2/\al c_{_F}\,\simeq\,(\chi u)^2\,(u\,\mp\,v)^2$
and the Eq.(56) should be used as far as $\Dl^2/\al c_{_F}$ can become zero,
while the main contribution is due to the region where $\Dl^2/\al c_{_F}$  
is strictly less than 1. So we obtain
\begin{equation}
{\cal F}_{\mp}\,\simeq\,-\,1\,+\,\pi^{1/2}\,\frac{\,\chi u\,(1-v^2)\,}
{|u\mp v|}\,.
\end{equation}
The part of $j$ which is due to the first term in (58) looks as follows:
\begin{equation}
-\,\frac{\,3\rho\,}\pi\,\frac{\,1\,}{\xi_0\chi^2}
\int\limits^{1/\chi}_0\,u\,{\text d}u\,
\int\limits^1_0\,\frac{{\text d}v}{\,(1-v^2)^{1/2}\,}\,=\,
-\,\frac{\,3\rho\,}4\,\frac{\,1\,}{\xi_0\chi^4}\,.
\end{equation}
The second term in (58) does not contribute at $v\ge 1/\chi$ because
sign$(u-v)=-1$, and therefore 
\begin{equation}
\frac{\,3\rho\,}{2\sqrt\pi}\,\frac{\,1\,}{\xi_0\chi}
\int\limits^{1/\chi}_0\,u\,{\text d}u\,
\int\limits^{1/\chi}_0\,{\text d}v\left({\text sign}(u\,-\,v)\,+\,1\right)\,=
\,\frac{\rho}{\,\sqrt\pi\,}\,\frac{\,1\,}{\xi_0\chi^4}\,.
\end{equation}
Then, the total contribution below $u=1/\chi$ is:
\begin{equation}
j^\prime\,=\,\frac{\,\rho\,}2\frac1{\,\xi_0\chi^4\,}\left(\frac2{\,\sqrt\pi\,}
\,-\,\frac{\,3\,}2\right)\,.
\end{equation}

At last, let us consider the rectangle $\{(u,v):\,$ $1/\chi\le u\le 1,
\,v\in[0,1]\}$. Here ${\cal F}_+$ can safely be expanded by the Laplace method.
As to ${\cal F}_-$, the integral diverges when $\Dl^2/\al c_{_F}$ falls into
the strip $|u-v|\le 1/\chi$ along the diagonale $u=v$ but the resulting
singularity is integrable. Here there is no interesting contribution as far as
$\x^2\rrr 1$ and the strip's width is $2/\chi$. Outside the strip the use of
(55) allows to fix unambiguously the logarithmic third order term. The 
experience of the previous calculations shows us that the coefficient at 
$(\xi_0\chi^4)^{-1}$ is mainly due to $0\le u\le 1/\chi$, and therefore its 
order of magnitude should be given by (61). So, we get with ${\cal F}_+$ and 
${\cal F}_-$:
$$
-\,\frac{\,\rho\,}{\pi}\,\frac{\,1\,}{\xi_0\chi^6}
\int\limits^1_{1/\chi}\,\frac{\,{\text d}u\,}{u^3}\frac{\,1\,}{(u^2+1)^{5/2}}\,
\int\limits^1_0\,\frac{{\text d}v}{\,(1-v^2)^{1/2}\,}
\left(u^2\,+\,8\,v^2\,-\,5\right)\,.
$$
The last equation results in the following contribution:
\begin{equation}
\frac{\,7\rho\,}8\,\frac{\,1\,}{\xi_0\chi^6}\,\log\frac{\,{\cal B}\,}
{\chi^2}\,,
\end{equation}
while the total quadratic correction (at least the order) is given by
\begin{equation}
\frac{\,\rho\,}2\frac1{\,\xi_0\chi^4\,}\left(\frac2{\,\sqrt\pi\,}
\,-\,\frac{\,3\,}2\right)\,\equiv\,\frac{\,\rho\,}2\,\frac{{\cal A}}
{\,\xi_0\chi^4\,}\,,\quad{\cal A}\,\approx\,\,-37\,\times10^{-2}\,.
\end{equation}

\section{DISCUSSION}

The present paper is concerned with the two main problems: to calculate the 
mass current $\vj$ in weakly inhomogeneous $\3$ using thermal Green functions 
and to obtain its asymptotic expansions at $T=0$ provided the London limit 
holds. Two main assumptions are of importance for our approach: the static 
order parameter can be linearized due to slowness of its spatial variation, 
and only those first order differentiations are retained in the chosen mixed
representation which are due to the kinetic energy of the $BCS-$Hamiltonian. 
Using slowness of the orbital vector texture we reduce the three initial 
dimensions to the one-dimensional situation so that the resulting operator 
${\cal H}_{em}$ has the simple form of the Hamiltonian of the Landau problem. 
Therefore, one can solve the governing Dyson--Gorkov equation exactly: just 
using the eigen-functions of ${\cal H}_{em}$ [22--24]. Thus a collection of 
exact formulas both for the Green function and $\vj$ appears which opens the
possibility to derive systematically the higher gradient corrections to the
dominant expression $\vj_0$ (1). The present paper completes the preceding
ones [22--24] which have been aimed at a more thorough resolution of the
mathematical difficulties of the Ref.[10]. Our approach unravels the situation
and provides a correct procedure to find the structure and the order of
magnitude of higher contributions to (1). The given approach is manifestly 
advantageous because the Laplace method is highly appropriate in the London 
limit.

Mathematically, we are mainly concerned with the $\xi$-integrated and then 
$\om$-summated normal Green function which results in two representations for 
$\vj$: the series and the integral. The integral one seems to be more 
attractive as far as it allows to obtain a self--contained expression (41) for 
the 
net correcting contribution. The last can satisfactory be studied by steepest 
descent due to the London limit holds, i.e. it can be expanded in powers of 
the orbital vector gradients. Particular limiting cases (zero temperature 
limit following the limit of lowest order in gradients, and {\it vice versa})
confirm the correctness of our manipulations. The known expression (1) 
arises as the lowest approximation without resorting to specific gauges for 
the order parameter. The appropriate ``commutative'' diagram is considered in 
the Sec.4.2. 

Three special orientations of $\rot\,\hl$ are considered in the Sec.5 to
deduce the correcting terms explicitly: $\rot\,\hl$ is parallel ({\it Example 
1}), and perpendicular ({\it Example 2}) to $\hl$, while the third {\it 
Example 3} implies an intermediate orientation of $\rot\,\hl$ with respect to 
$\hl$. Corrections are considered up to third order in gradients of $\hl$ and 
they are not only of pure polynomial type. Namely, new cubic corrections are 
found which contain the logarithm of the London parameter. By comparison with 
[10], we provide the numerical coefficients at the second order terms.
The nodes vicinities
$0\le\theta$ (or $\pi -\theta$) $\lll 1/\chi$ on the Fermi sphere give the
dominant contribution to the numerical coefficients at the pure power terms.

In the first case only the third order logarithmic correction is present. In 
the second 
case both the lowest ones appear: the quadratic and the cubic. As it is clear 
from the analysis [10], all the quadratic corrections should be proportional 
to $\lrot$, and that correlates with the absence of quadratic term in 
{\it Example 1}. We have specified the {\it Example 2} so that $\vj_
{{\text{corr}}}$ is orthogonal to $\hl$. Thus the second order term in (51)
would correspond to that which has been written in [10] in the form 
$(\rot\,\hl)_{\perp}\lrot$, and the corresponding numerical coefficient is 
${\cal A}$ (52) [23]. {\it Example 3} also results in the both lowest 
corrections: the quadratic (63) and the cubic (62). As to the quadratic 
correction along $\hl$, the answer of [10] reads:
$$
\rho\,\xi_0\,\lrot\left(\,A\,(v_3\,-\,\hl\cdot\rot\,\hl/4m)\,+\,
B\,(\cd_1\hl_2\,+\,\cd_2\hl_1)\right)\,.
$$
The term (63) above is just to be compared with the last
expression at $A\ne 0$, $B=0$. It is clear from (36) that the result obtained 
by us is also applicable to establish the contribution at $A=0$, $B\ne 0$.

Moreover, our situation is more rich than in [10] because the logarithmic 
corrections are demonstrated. Indeed, let us recall the correction found in 
[13]. As it is seen from (36), the components of the superfluid velocity 
$m\vec v$ and gradients of $\hl$ enter equally to the parameter $\al$, and 
therefore $\chi_{{\text{orb}}}\,(\vec v_s\cdot\vep)\,\wh l_a\vep\,\,\wh l_a$, 
where $\chi_{{\text{orb}}}$ is 
logarithmically large [9], should be treated as the logarithmic third order 
term, and it would correspond to the third order contribution in (62). 

To conclude, the investigation presented which is based on thermal Green 
functions demonstrates various corrections of second and third order to the 
mass current (1) at $T=0$ provided the London limit condition holds.
The corrections predicted early in [10,~13] can systematically be
deduced in our approach. The representations obtained for $\vj$ and 
$\vj_{{\text{corr}}}$ would serve a basis for further investigations.

\medskip
\subsection*{ACKNOWLEDGEMENTS}
\medskip

The author would like to express his gratitude to F.Gaitan, A.G.Izergin, 
N.B.Kopnin, M.Krusius, E.Thuneberg and G.E.Volovik for illuminating 
discussions. It is pleasure to thank the Low Temperature Laboratory (Helsinki 
University for Technology) for warm hospitality. 
The research described in this publication was made possible in part by the 
Grant No R4T300 from the International Science Foundation and Russian
Government, and also by the Russian Foundation for Fundamental Research 
Projects No 94--02--03712 and No 96--01--00807.

\medskip
\section*{APPENDIX I}
\medskip

The following Mehler formula is valid:
$$
\sum\limits_{n=0}^\infty\,a^n\psi^2_n(y)\,=\,\frac1{\sqrt{\pi(1-a^2)}}\,
\exp\left(-\,\frac{1-a}{1+a}\,y^2\right)\,,\qquad|a|\,<\,1\,,
\eqno({\text AI}.1)
$$
where $\psi_n(y)$ are the Chebyshev--Hermite functions (see generating 
functions of Hermite polynomials in [28]). Using
$$
(n\,+\,q)^{-1}\,=\,\int\limits^\infty_0\,{\text d}t\,e^{-t(n+q)}\,,\qquad
n\,\ge\,0\,,\,q\,>\,0\,,
$$
and (AI.1), we obtain:
$$
\sum\limits^\infty_{n=0}\,\frac{\psi^2_n(y)}{n+q}\,=\,
\frac1{\sqrt{2\pi\,}}\,\int\limits^\infty_0\,\frac{{\text d}t}
{\sqrt{\sinh t\,}}\,\exp\left(\left(\frac{\,1\,}2-q\right)t\,-\,y^2\tanh(t/2)
\right)\eqno({\text AI}.2)
$$[22].

\medskip
\section*{APPENDIX II}
\medskip

Here we obtain $\al$ (35) as in [10]. To this end let us rewrite 
$$
\Dl(\hk,\r)\,=\,\dl (\hk\cdot\widehat\Dl_1(\r) +i\hk\cdot\widehat\Dl_2
(\r))
\eqno({\text AII}.1)
$$
with the help of
$$
\widehat\Dl_{\al}(\r)\approx\widehat\Dl_0({\cal O})+\r\cdot\vep\,\widehat\Dl_
{\al}({\cal O})\,,\quad\quad(\al=1,2),
\eqno({\text AII}.2)
$$
where the derivatives $\partial_i\widehat\Dl_{\al}({\cal O})$ are 
linear functions of $\widehat\Dl_{\al}({\cal O}):$
$$
\partial_i\widehat\Dl_{\al}({\cal O})\,=\,\vec \om^{i}\times
\widehat\Dl_{\al}({\cal O}).
\eqno({\text AII}.3)
$$
The Eq.(AII.3) acquires the form of the Mermin--Ho relation [30] provided
the identifications
$$
\om^i_3\,=\,-2mv_i\,,\qquad\om^i_2\,=\,\partial_i\hl_1\,,\qquad\om^i_1\,=\,-\partial
_i\hl_2
\eqno({\text AII}.4)
$$
are made. Substituting (AII.2)--(AII.4) to (AII.1) one gets
$$
\Dl(\hk,\r)\,=\,\dl (\hk_1\,+\,i\hk_2)\,+\,
\rho\,i\dl\,\left(2mv_p\hk_p(\hk_1+i\hk_2)-
\widehat k_3\widehat k_p(\cd_p\wh l_2-i\cd_p\wh l_1)\right)\,\equiv
$$
$$\equiv\,\dl\,\sin\theta\,e^{i\phi}\,+\,\rho\left[\dl\left(...\right)\,e^{i(
\pi/2-\psi)}\right]\,e^{i\psi}
\eqno({\text AII}.5)
$$
Eventually, the square brackets in (AII.5) are denoted as $\al$, and the 
factor $e^{i\psi}$ is to make it a real positive fixing therefore 
$\widehat\Dl_1\,,\,\widehat\Dl_2$ in the plane perpendicular to $\hl$. From 
(AII.5) it is seen that $\al c_F$ is a linear form of gradients which can be 
written formally as
$$
\al c_F\,=\,\dl^2\xi_0 \sum\,\left({\text gradients}\right)\,=
\,\frac{\dl^2}{\chi^2}\sum\,\frac{{\text gradients}}{|{\text gradients}|}.
\eqno({\text AII}.6)
$$
By (8) we consider $\al c_F/\dl^2$ as small parameter.

\end{document}